\documentclass[12pt,preprint]{aastex}

\renewcommand\email\texttt

\def\spose#1{\hbox to 0pt{#1\hss}}
\def\lta{\mathrel{\spose{\lower 3pt\hbox{$\sim$}}
    \raise 2.0pt\hbox{$<$}}}
\def\gta{\mathrel{\spose{\lower 3pt\hbox{$\sim$}}
    \raise 2.0pt\hbox{$>$}}}
\newcommand{\boo}{Bo\"otes\,II}
\begin{document}


\shorttitle{\sc A Pair of Bo{\"o}tes} \shortauthors{Walsh, Jerjen
and Willman}

\title{A Pair of Bo{\"o}tes: A New Milky Way Satellite}

\author{S.\ M. Walsh\altaffilmark{1},
H. Jerjen\altaffilmark{1},  B. Willman\altaffilmark{2}}

\altaffiltext{1}{Research School of Astronomy and Astrophysics,
Australian National University, Cotter Road, Weston, ACT 2611,
\email{swalsh, jerjen@mso.anu.edu.au}} \altaffiltext{2}{Clay Fellow,
Harvard-Smithsonian Center for Astrophysics, 60 Garden Street
Cambridge, MA 02138}


\begin{abstract}
As part of preparations for a southern sky search for faint Milky
Way dwarf galaxy satellites, we report the discovery of a stellar
overdensity in the Sloan Digital Sky Survey Data Release 5, lying at
an angular distance of only $1.5$\,degrees from the recently
discovered Bo{\"o}tes dwarf. The overdensity was detected well above
statistical noise by employing a sophisticated data mining algorithm
and does not correspond to any catalogued object. Overlaid
isochrones using stellar population synthesis models show that the
color-magnitude diagram of that region has the signature of an old
(12\,Gyr), metal-poor (${\rm Fe/H}\approx-2.0$) stellar population
at a tentative distance of $60$\,kpc, evidently the same
heliocentric distance as the Bo\"otes dwarf. We estimate the new
object to have a total magnitude of $M_{V}\sim-3.1\pm1.1$\,mag and a
half-light radius of $r_{h}=4'.1\pm1'.6$ ($72\pm28$\,pc) placing it
in an apparent $40<r_{h}<100$\,pc void between globular clusters and dwarf galaxies, occupied only by another recently
discovered Milky Way Satellite, Coma Berenices.
\end{abstract}



\section{Introduction}
The last three years have seen a torrent of new Milky Way (MW)
satellites being discovered in the Northern hemisphere, almost
doubling the number known prior to 2005: Bo\"otes (Belokurov et
al.~2006a), Canes Venatici (Zucker et al.~2006a), Willman 1 (Willman
er al.~2005a), Ursa Major (Willman et al.~2005b), Ursa Major II
(Zucker et al.~2006a), Hercules, Coma Berenices, SEGUE 1, Canes
Venatici II, Leo IV (Belokurov et al.~2006b), and Leo T (Irwin et
al.~2007). Eight of these new objects are consistent in size and
luminosity with dwarf spheroidal satellites while Willman 1 and
Segue 1 straddle the intersection of dwarfs and globular clusters.
Coma Berenices falls in an apparent void of objects spanning the
$40-100$\,pc range (see Fig 1. Gilmore et al.~2007).
These objects were all initially detected as overdensities of
resolved stars in the photometric data of Sloan Digital Sky Survey
(SDSS, York et al.~2000) and, with the exception of Leo IV,
subsequently confirmed with follow-up observations. The numerous
discoveries of extremely low surface brightness dwarf spheroidals
(dSphs) in the ~1/4 of the sky covered by SDSS strongly suggests
that there are many more yet undiscovered. A significant new
population of dwarf satellite galaxies would go a long way to
reconcile the current discrepancy between Lambda Cold Dark Matter
theory predictions (Klypin et al.~1999; Moore et al.~1999) and
actual observed dSph numbers.

The next few years will see the advent of digital surveys that will
enable all-sky searches for Milky Way satellites (e.g. PanSTARRS,
Kaiser et al.~2005).  We intend to blindly scan the entire Southern
sky (20,000 square degrees) for new MW dwarf satellites with the
upcoming Southern Sky Survey performed with the 1.3 meter ANU
SkyMapper telescope at Siding Spring (Keller et al.~2007). The final
combined $\sim25$\,TB catalog of the survey is estimated to reach a
signal-to-noise of 5 at $r=22.6$, 1.0 mag deeper than SDSS. In
preparation for this survey we are testing sensitive data mining
algorithms and search strategies using the freely available SDSS
Data Release 5 (DR5, Adelman-McCarthy et al.~2006). A full overview,
results and a detailed discussion will be presented in a subsequent
paper (Walsh et al. in prep). Our software test on SDSS data has
yielded several promising candidates, the most prominent of which we
present here. While follow-up observations will reveal the
candidate's true nature in more detail, its size and luminosity are
consistent with those of the other recent detections that have been
labeled dwarfs spheroidals. We thus follow convention and designate
it Bo\"{o}tes II.

\section{Data and Discovery}

DR5 includes a five color photometric catalogue covering 8000 square
degrees around the north Galactic pole (Adelman-McCarthy et
al.~2006). We have searched this publicly available data for
concentrations of old stars at various distance intervals out to the
Galactic virial radius of 250\,kpc. We use a method similar to that
described in Willman et al. (2002) and Willman (2003), and described
in full in Walsh et al. (in preparation). We use a complicated set of cuts to identify all stellar sources in
fields of $3^{\circ}$ height in Declination and of arbitrary width
in Right Ascension that are consistent in $(g-r,r)$ parameter space
with that of a dSph at a desired distance (Red Giant Branch, Blue Horizontal Branch and Main-Sequence Turnoff). We then convolve the
binned spatial positions of these sources with an exponential surface
brightness profile and subtract the $0.9^{\circ}\times0.9^{\circ}$ running mean from each $0.02^{\circ}\times0.02^{\circ}$ pixel. A density
threshold in standard deviations above the local mean is defined as a function of the
background stellar density for each pixel, allowing us to search
over fields with stellar density gradients. The process is repeated
for different magnitude bins to change sensitivity with distance.

Applying this method to DR5 we recovered all of the recently
reported dSphs, as well as many previously known objects such as
globular clusters and background galaxy clusters. The detection
significance of the faintest dwarfs, quantified by the parameter
$P$ (maximum level above threshold density times area above threshold), are shown in Fig.\,\ref{fig:hist}, along with the result for
the new object. The solid line shows the number of ``detections'' in
thirty-nine 1000 square degree randomized stellar fields each of
varying stellar density to determine foreground contamination from
random clustering. The newly discovered satellites (minus SEGUE 1
which fell outside the analyzed area)  as well as the \boo\,
overdensity are all well above the threshold $P=85$ above which our
detection algorithm statistically yields less than one false
positive detection over the entire area of DR5. The Bo\"otes II
overdensity is not associated with any known Galactic or
extragalactic object and is consistent in $(g-r,r)$ and
size-luminosity space with a new dwarf. Figure \ref{fig:pair} shows
the position of Bo\"otes\,II relative to Bo\"{o}tes.


Fig \ref{fig:boo2} shows our detections of Coma Berenices, \boo\,
and Bo\"otes. The contours represent the level above the threshold
density, which is then multiplied by the detection area to give $P$.
Coma Berenices and \boo\,peak at higher densities because they are
more concentrated than Bo\"otes, but the latter's spatial extent
means that it is still a stronger detection. The detection of \boo\,
is consistent in all respects with the detections of the other
Galactic satellites, albeit much fainter.

\section{Candidate Properties}
We use SDSS data to extract as much information as possible and estimate preliminary parameters for \boo. The overdensity is
visible even before smoothing and the lack of a concentration of
background galaxies (Figure \ref{fig:fits}) allows to exclude a
galaxy cluster as an origin. Looking at the CMD in Figure
\ref{fig:boo2} reveals a weak red giant branch and blue horizontal
branch (or red clump) at a distance modulus apparently identical to
that of Bo\"{o}tes, and similar to the Coma Berenices dwarf
($m-M=18.2$, Belokurov et al.~2006b). The CMD features become even
more prominent in the associated area-normalized field-subtracted
Hess diagram. Overplotted is the isochrone of a metal-poor
([Fe/H]$=-2.0$), old (12\,Gyr) stellar population (Girardi et
al.~2004) to illustrate the consistency of our object with an old
stellar population. Using the assumed distance modulus of
$(m-M)=18.9$ (60\,kpc) which is the same heliocentric distance of
the Bo\"otes dwarf, \boo\, would lie at a spatial distance of only
$\sim1.6$\,kpc from Bo\"{o}tes. This hints at a physical connection
between the two systems, although a further discussion of this idea
is beyond the scope of this letter.

Figure \ref{fig:sb} presents the azimuthally averaged stellar
density profile generated from all stars with $(g-r)<0.65$ and
$17.0<r<22.5$ centered on \boo. These data were then fitted with a
Plummer profile (dotted line) plus a constant (dashed line), the
latter to account for the foreground screen of Galactic stars. The
best-fitting profile has a half-light radius of $4.1\pm1.6$\,arcmin,
approximately one third of the physical size of Bo\"{o}tes.
Alternatively, fitting an exponential profile to the data gives a
half-light radius of $4.0\pm1.9$\,arcmin.

We use two methods to empirically derive the total magnitude of our
object. Firstly we use SDSS coverage of the Draco dSph to calculate
the flux ratio of the integrated luminosity functions. We derive a
flux ratio $f_{Draco}/f_{Boo\,II}\simeq 172\pm38$. This converts
into a magnitude difference of $2.1\pm0.3$\,mag and a total absolute
magnitude of $M_{V}\sim -3.8\pm0.6$\,mag for \boo\,, adopting
$M_{V}=-9.4$\,mag for Draco (Grebel et al.~(2003)). The same
analysis yields $M_{V}\sim -6.0\pm0.6$ for Bo\"{o}tes, 0.7\,mag
brighter than $M_{V}\sim -5.3\pm0.6$ given by Belokurov et
al.~(2006a). Secondly, we use the integrated surface brightness
profiles of Bo\"otes and \boo. The flux ratio from this method gives
$f_{Boo}/f_{Boo\,II}\simeq 15$. Using $M_{V}\sim -5.3\pm0.6$ for
Bo\"otes gives $M_{V}\sim -2.4\pm0.6$. We therefore adopt a result
of $M_{V}\sim -3.1\pm1.1$.

\section{Discussion and Conclusion}
\begin{figure}
\plotone{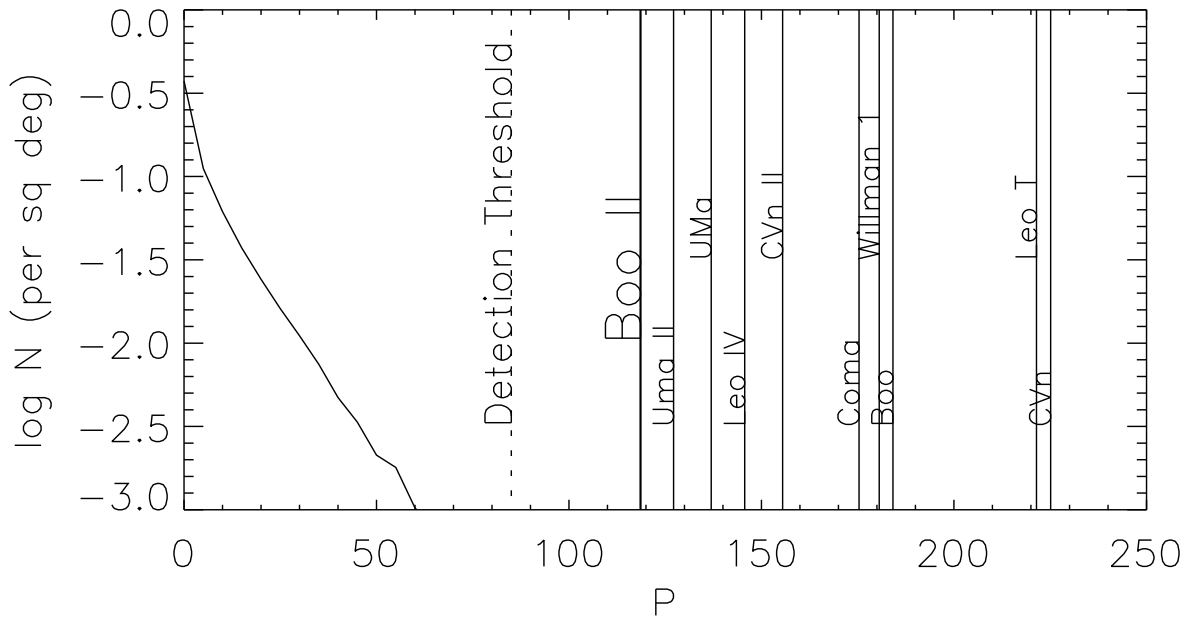} 
\caption{Curve of false positive detections (black
line) as a function of the detection parameter $P$ provided by our
data mining algorithm. The detection threshold (vertical dashed
line) is located at $P=85$ above which our detection algorithm
statistically yields less than one false detection over the entire
8800 square degrees of SDSS-DR5. Overplotted are the detection
levels for the faintest known dSphs as well as the Bo\"otes II
object. With $P=119$, the Bo\"otes II overdensity is detected just
below the UMa II ($P=128$) and UMa ($P=137$) dwarfs.
\label{fig:hist}}
\end{figure}

\begin{figure}
\epsscale{.70}
\plotone{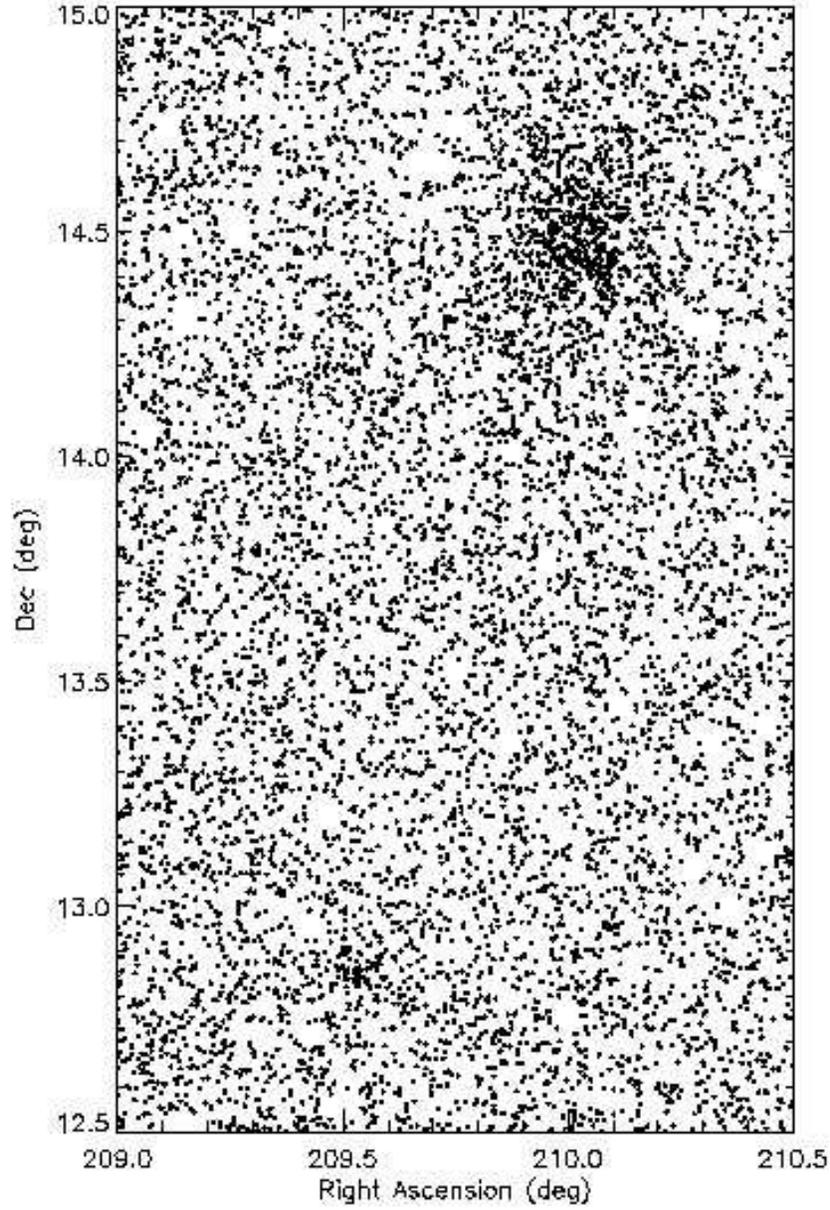} 
\caption{Positions of all SDSS stars with
$17<r<23$\,mag and $g-r<0.65$. The Bootes dwarf is clearly visible
(210.05d +14.50) as is our candidate $1.5$\,degrees to the
south-west at 209.55d +12.85.} \label{fig:pair}
\end{figure}

\begin{figure}
\epsscale{1.0}
\plotone{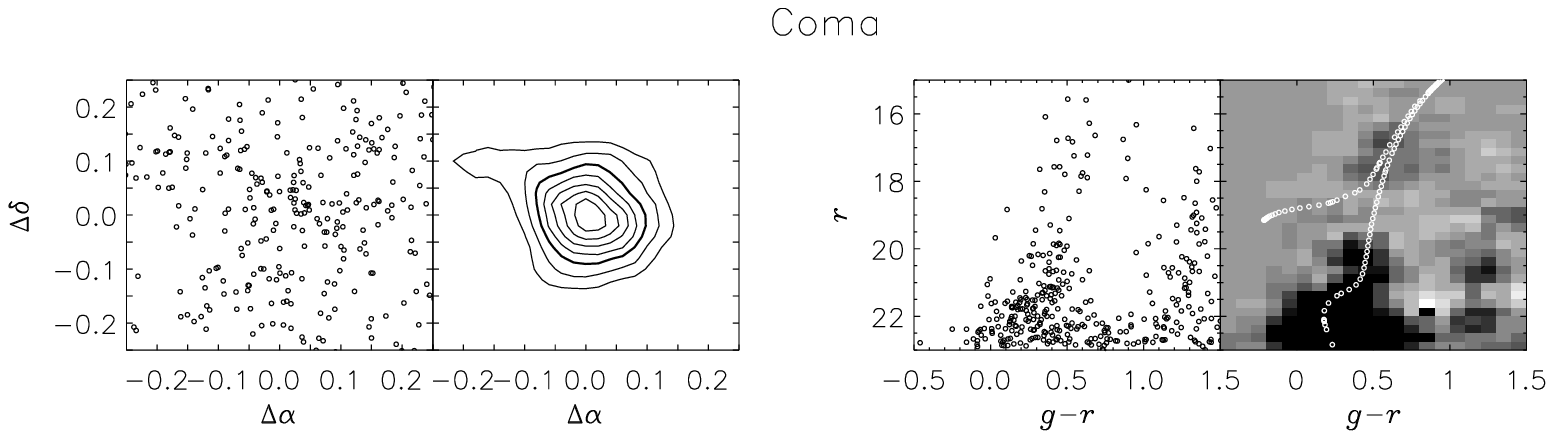} 
\plotone{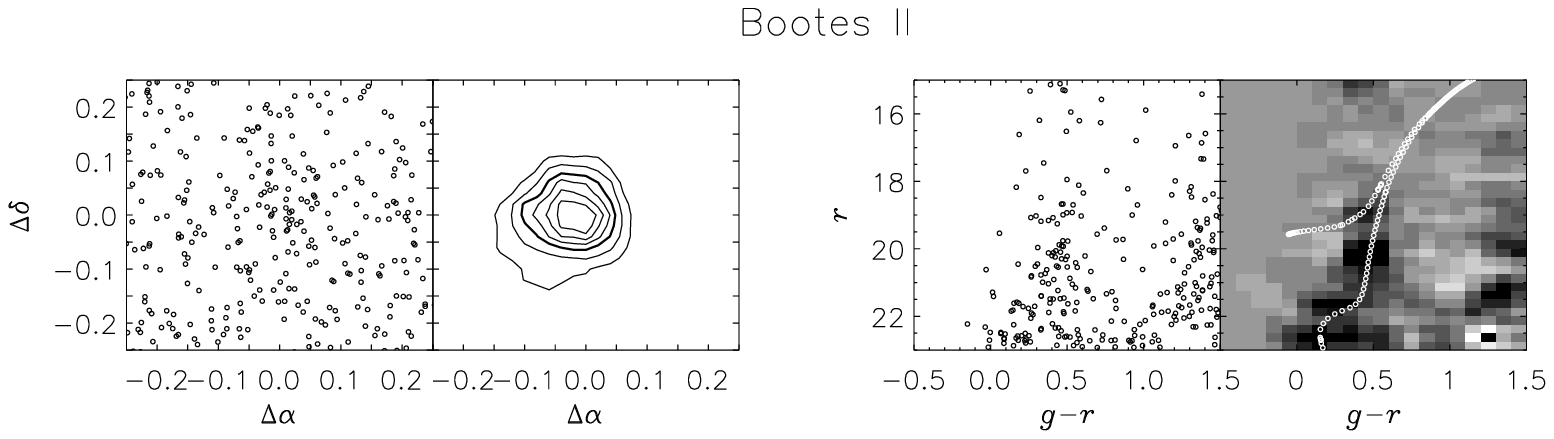} 
\plotone{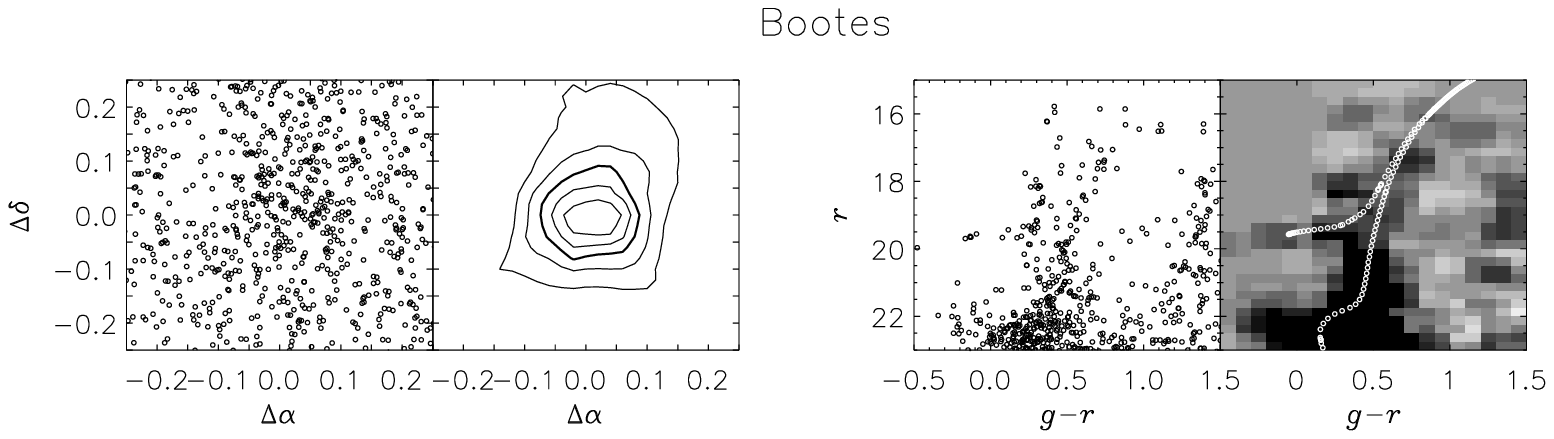}
\caption{From top to bottom: Coma Berenices, \boo, Bo\"otes. {\em Left Panels}:
positions of SDSS stars passing the photometric selection criteria.
{\em Middle Left}: Smoothed positions with contours at 0.5,0.75,1.0
(thick line),1.2,1.4,1.6 and 1.8 multiples of threshold density.
{\em Middle Right}: CMD of region within the 1.0 contour. {\em Right
Panels}: field subtracted Hess diagrams of same regions with
overlayed stellar isochrone.\label{fig:boo2}}
\end{figure}

\begin{figure}
\plotone{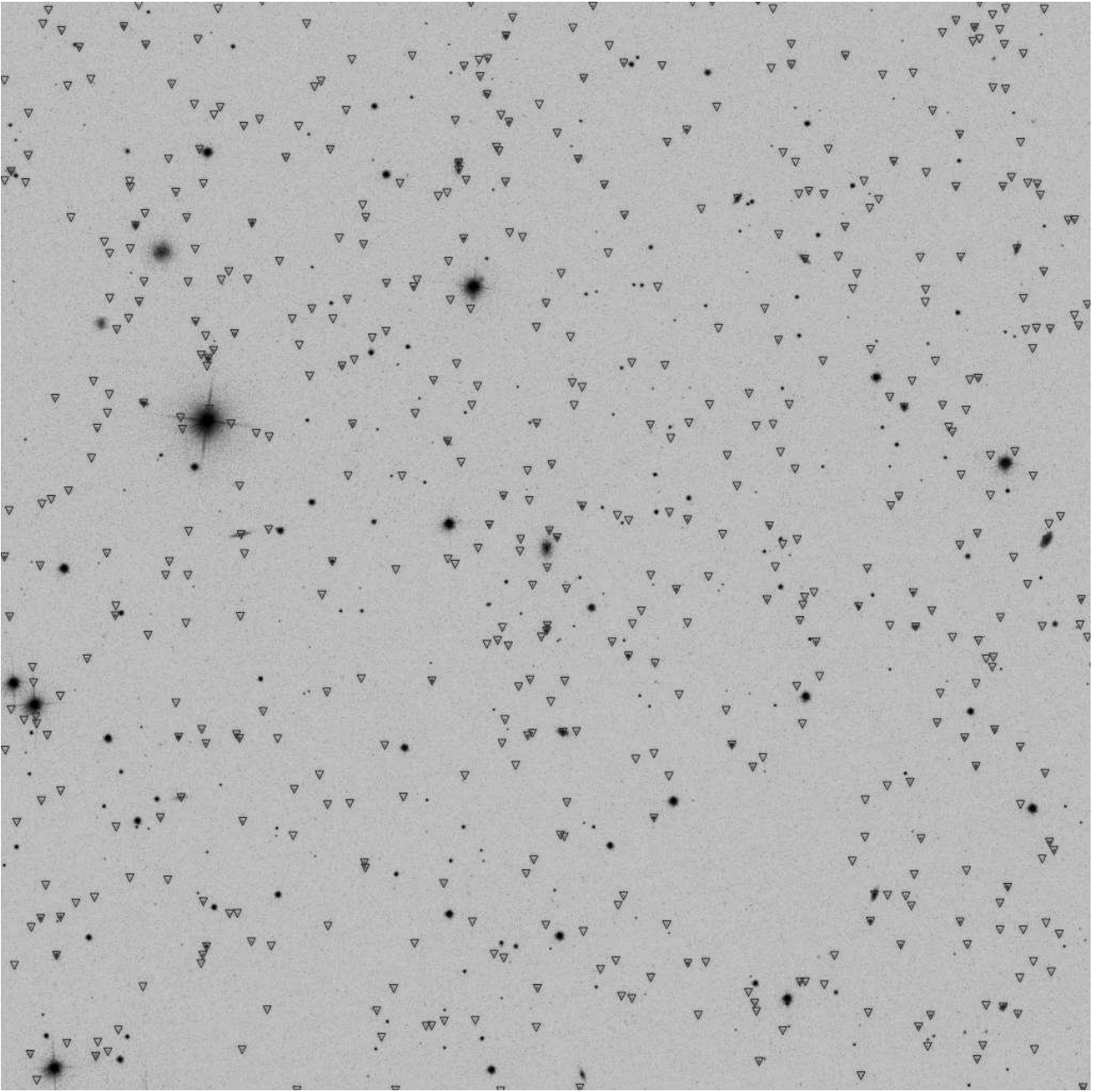}
\caption{SDSS image ($15\prime\times15\prime$) centered on the detection of
 Bootes II with SDSS galaxies overlayed (triangles). Left is East and Up is North.}
\label{fig:fits}
\end{figure}

\begin{figure}
\plotone{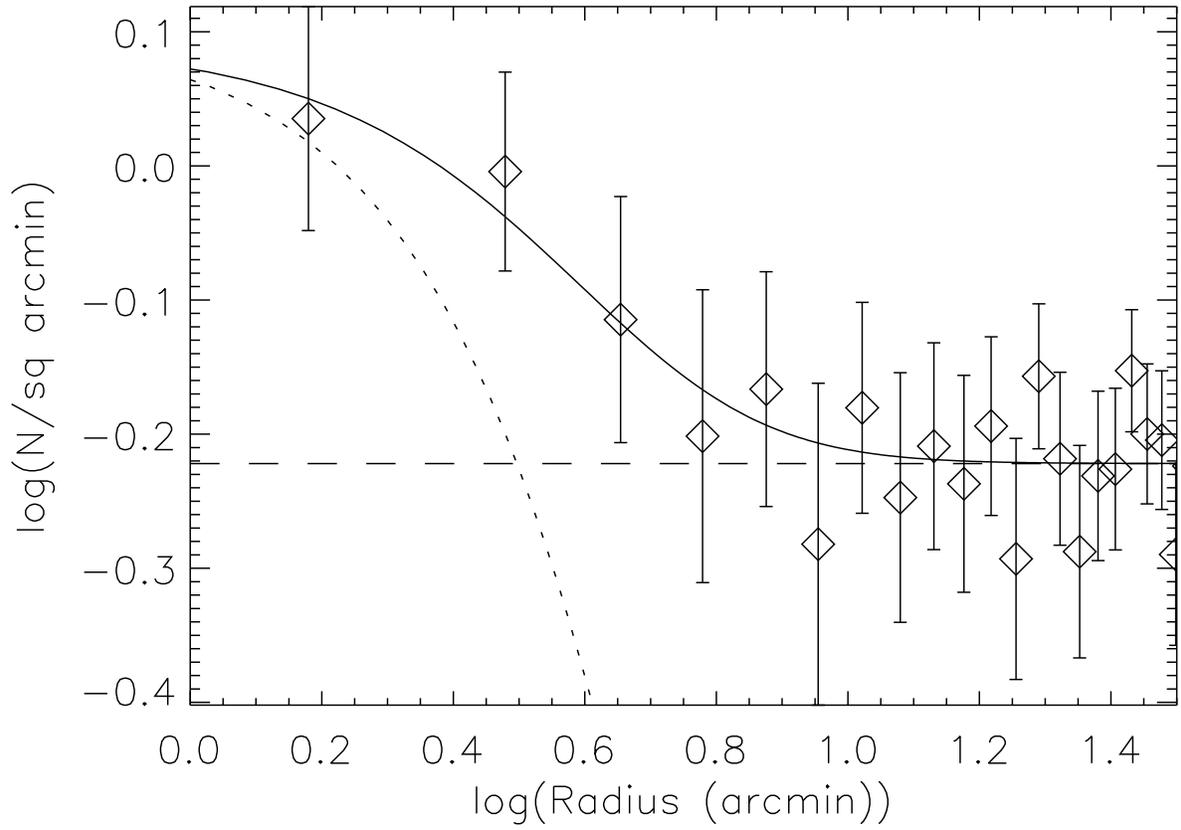}
\caption{Radial stellar density profile generated from all stars
 with $(g-r)<0.65$ and $17.0<r<22.5$ centered on the new object (diamonds).
 The dotted line is a Plummer profile with a scale parameter $a=5.3$\,arcsec and
 the dashed line is the contribution of foreground stars. The solid line is the combined fit.}
\label{fig:sb}
\end{figure}

We report a new Galactic satellite called \boo, only $\sim1.5$
degrees away from the Bo\"{o}tes dwarf. This object was discovered
as a resolved stellar overdensity in an automated search of SDSS
DR5. Any object that is detected by our algorithm will fall in one
of the following categories: random foreground clustering, galaxy
clusters, stellar associations of partially resolved nearby
galaxies, globular clusters, or Galactic dwarf spheroidals. Random
clustering at this level is extremely unlikely with $\sim0.008$ such false objects occuring in the entire DR5 area (Fig.~\ref{fig:hist}). The CMD shows an apparent MST and RGB structure that is unlikely to be associated with a distant galaxy cluster and no
evidence is found of a suspicious accumulation of background
galaxies (Fig.~\ref{fig:fits}).

\begin{figure}
\plotone{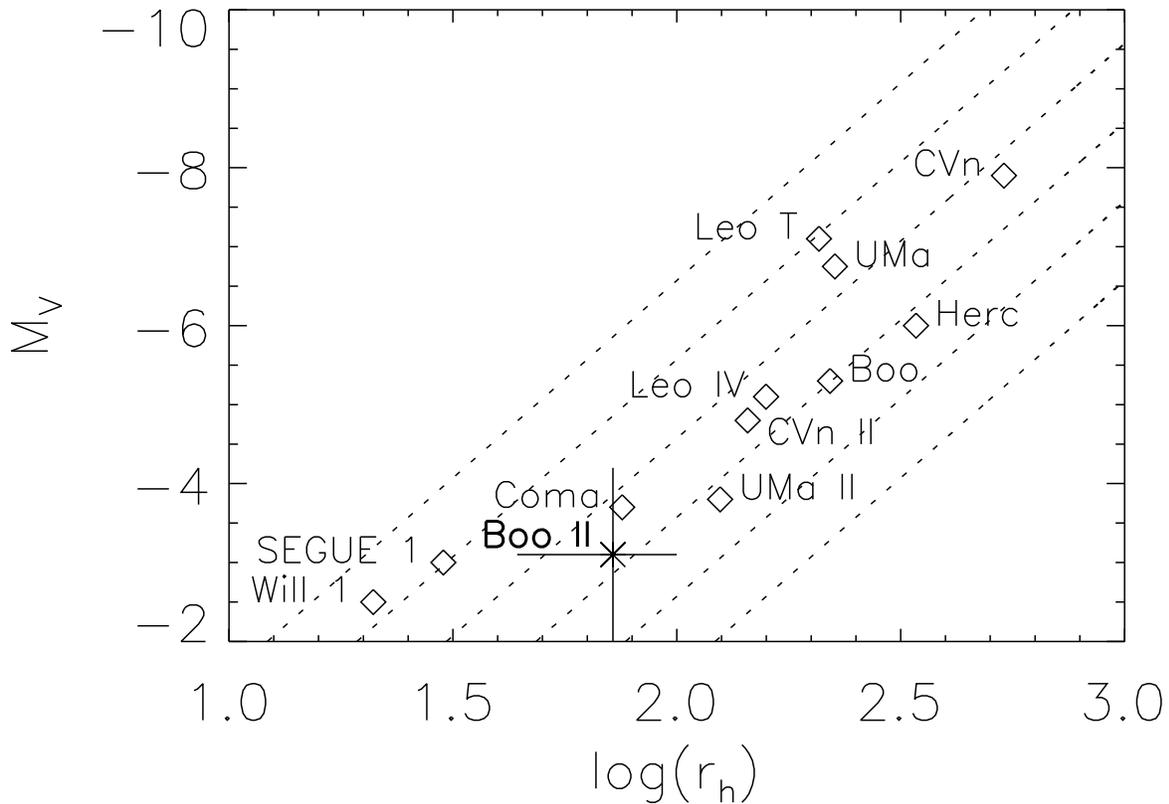}
\caption{Absolute magnitude ($M_{V}$) versus half-light radius
 ($r_h$ in pc) for all Galactic companions discovered in SDSS.
 Bo\"otes II is shown as an asterisk, with associated error bars.
 Lines of constant central surface brightness are shown with
 $\mu_{V}=27, 28, 29, 30, 31$ and $32$\,mag\,arcsec$^{-2}$ from left to right.}
\label{fig:lr}
\end{figure}

The combined evidences from CMD, surface brightness profile and the
good agreement with the size-luminosity relationship of other MW
satellites leads us to conclude this object is a previously
undiscovered companion to the Milky Way at a tentative distance of
$60$\,kpc. But is it a GC or a dSph? Traditionally, these objects
are distinguished by their differing physical size; dark matter
dominated dSphs are more extended than a purely stellar system of
equal luminosity. Equipped with the two parameters $\log(r_h/{\rm
pc})=1.8^{+0.2}_{-0.3}$ and $M_{V}=-3.1\pm1.1$\,mag for \boo\, we
add our object to the other recently discovered dwarfs in the
size-luminosity plot (Fig.~\ref{fig:lr}). The distinction between
dSphs and GCs is blurred in the low luminosity regime as was
emphasized by the discovery of Willman 1 (Willman et al.~2005a).
\boo\, falls alongside Coma Berenices in the 40--100\,pc region devoid of other objects, between globular clusters and dwarfs.
\boo\, is of comparable luminosity to SEGUE 1, but is a factor of
$\sim2$ larger.

With a physical half-light size that is an order of mag larger than
most GCs but similar to those of dSphs, we are inclined to designate
\boo\, a dwarf galaxy.  However, without kinematic data, it is not
possible to say for certain whether or not \boo\, formed inside of a
dark matter halo, which would confirm it as such. It is also not
possible with the current data to determine the extent to which
tidal effects may have shaped the observed size and luminosity of
\boo. A combination of follow-up spectroscopy and deep imaging will
not only enable a more robust evaluation of the dark matter content
of this object, but will also enable an evaluation of a possible
physical relationship between Bo\"otes and \boo.

\acknowledgments We thank Pavel Kroupa, Manuel Metz and Jose Robles
for useful discussions and comments in preparation of the
manuscript.

Funding for the SDSS and SDSS-II has been provided by the Alfred P.
Sloan Foundation, the Participating Institutions, the National Science
Foundation, the U.S. Department of Energy, the National Aeronautics and
Space Administration, the Japanese Monbukagakusho, the Max Planck
Society, and the Higher Education Funding Council for England. The SDSS


\clearpage

\begin{deluxetable}{lrlr}
\tablewidth{0pt} 
\tablecaption{\label{tbl:pars}} 
\tablehead{
\colhead{Parameter} & 
\colhead{Bo\"{o}tes II}
}
\startdata RA (h m s) &13 58 00 & {[Fe/H]} & $-2.0$ \\
Dec (d m s)&+12 51 00 & Age (Gyr) & 12 \\
 $(l,b)$ & (353.7,68.9) & $\mu_{0,\rm V}$ (Plummer)&  $29.8\pm0.8$ \\
 $(m-M)$& $18.9\pm0.5$ & $r_{h}$ (Plummer) & $4'.1\pm1'.6$\\
Distance (kpc) & $60\pm10$ & $\mu_{0,\rm V}$ (Exponential)& $29.6\pm0.8$ \\
$M_{V}$ (mag) & $-3.1\pm1.1$&$r_{h}$
(Exponential)&$4'.0\pm1'.9$
\enddata
\end{deluxetable}

\clearpage

\clearpage

\clearpage

\clearpage

\clearpage

\clearpage

\end{document}